\documentclass[prl,twocolumn]{revtex4}

\newcommand{\cddyz}{Cd$-d_{yz}$}
\newcommand{\cgav}{V$_{33}$}

\newcommand{\neutra}{\mbox{Cd$^0$}}
\newcommand{\carg}{\mbox{Cd$^-$(2e)}}
\newcommand{\fluor}{\mbox{Cd$^-$(fluorine)}}
\newcommand{\unitefg}{\mbox{10$^{21}$V/m$^2$}}
\usepackage{dcolumn}
\usepackage{color}
\usepackage{graphicx}

\begin{document}

\title{{\bf Anisotropic relaxations introduced by Cd impurities
in Rutile TiO$_2$: first-principles calculations and experimental
support}}
\author{L.A. Errico, G. Fabricius, and M. Renter\'{\i}a}
\altaffiliation[Also at ]{LURE, Universit\'e Paris-Sud, Orsay,
France.} \affiliation{Departamento de F\'{\i}sica, Facultad de
Ciencias Exactas, Universidad Nacional de La Plata, C.C.
N$^\circ$67, 1900 La Plata, Argentina}
\author{P. de la Presa and M. Forker}
\affiliation{Institut f\"ur Strahlen- und Kernphysik,
Universit\"at Bonn, Nussallee 14-16, 53115 Bonn, Germany }
\date{\today}

\begin{abstract}

We present an {\it ab initio} study of the relaxations introduced
in TiO$_2$ when a Cd impurity substitutes a Ti atom and the
experimental test of this calculation by a
perturbed-angular-correlation (PAC) measurement of the orientation
 of the electric-field gradient (EFG) tensor at the Cd site.
 The {\it ab-initio} calculation predicts strong anisotropic
relaxations of the nearest oxygen neighbors of the impurity and a
change of the orientation of the largest EFG tensor component,
V$_{33}$, from the $[001]$ to the $[110]$ direction upon
substitution of a Ti atom by a Cd impurity. The last prediction is
confirmed by the PAC experiment that shows that V$_{33}$ at the Cd
site is parallel
 to either the $[110]$ or  the $[1\overline{1}0]$ crystal axis.

\end{abstract}

\maketitle

Metal impurities in oxides are a challenging problem not only from
a fundamental point of view. A reliable calculation of impurity
properties such as, e.g., their energy levels in semiconductor
oxides is also of great technological importance. Impurities
introduce structural atomic relaxations in the host and modify the
electronic structure of the system and  it is the interplay of
these two effects which makes an adequate theoretical description
very difficult. Recently, there have been several attempts to deal
with this problem at the {\it ab initio} level
\cite{impurities,cdte_blaha}.

Experimental tests of the {\it ab initio} predictions are of
fundamental importance for the evaluation of models of
impurity-induced relaxations. The structural relaxations modify
the local electron distribution at the impurity site. A quantity
reflecting this distribution is the electric-field gradient (EFG).
As the EFG decreases with 1/r$^3$ from the producing charge
density, it is very "short-sighted" and therefore particularly
sensitive to slight local changes. Consequently, the EFG at the
impurity site carries information on the relaxations induced by
the impurity and its experimental value may serve as test of the
{\it ab initio} calculations. The EFG can be determined by a
measurement of its electric hyperfine interaction with  a
quadrupole moment of the impurity nucleus.

In this paper we present an {\it ab initio} calculation of the
relaxations at Cd impurities on substitutional Ti sites in TiO$_2$
and  the experimental test of this calculation by a measurement of
the EFG tensor experienced by the Cd impurities. This particular
host (TiO$_2$)-impurity (Cd) combination was suggested mainly by
two aspects. First, strength, symmetry, and orientation of the EFG
tensor at Ti nuclei in the rutile structure of  pure TiO$_2$ are
known from previous NMR measurements of the $^{49}$Ti quadrupole
interaction in single crystals \cite{nmr_tio2}. {\it Ab initio}
calculations \cite{blaha_tio2} are in agreement with this
experimental result. Second,
 $^{111}$Cd is one of the most favorable probe nuclei for
perturbed-angular-correlation (PAC) experiments and when
introduced as an impurity into TiO$_2$ is known to reside on
substitutional Ti sites. PAC is a hyperfine spectroscopic
technique that can be applied at very low probe concentrations and
is therefore particularly well suited for the investigation of the
EFG at isolated impurities. Strength (in terms of the absolute
value of the largest principal-axis component, \cgav) and symmetry
of the EFG tensor of $^{111}$Cd on Ti sites in TiO$_2$ have been
measured previously by $^{111}$Cd PAC in polycrystalline TiO$_2$
\cite{lieb,catchen}.

 Recently, the EFG of Cd in TiO$_2$ has been calculated by Sato {\it et al.} \cite{akai}
imposing the constraint of isotropic  relaxations of the nearest
oxygen neighbors and by some of us  without this constraint
\cite{our_nqi}. Although the absolute value of V$_{33}$ provided
by experiment is fairly well reproduced, the results of both
calculations differ considerably: $(i)$ opposite signs are
predicted and $(ii)$ according to our calculation the substitution
of Ti by Cd should change the EFG orientation from the $[001]$
direction of pure TiO$_2$ to the $[110]$ direction while according
to Sato {\it et al.} the $[001]$ direction is not affected. These
two approaches to the problem of a Cd-impurity in TiO$_2$ have two
shortcomings: $(i)$ the size of the supercell (SC) considered in
both calculations (12 atoms) was very small for a dilute impurity
calculation and $(ii)$ the charge state of the impurity was not
taken into account in a self-consistent way. In the present work
we have removed both shortcomings by using much bigger SCs and
considering different impurity charge states in a fully
self-consistent way.

\begin{figure}[!]
\includegraphics*[bb= 3.2cm 20.0cm 10cm 10cm, viewport=-0.7cm 0cm 8.6cm 6cm, scale=0.75]{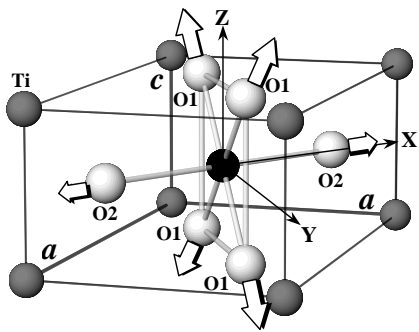}
\caption{\label{fig1} Unit cell of rutile TiO$_2$. All the results
discussed in this article refer to the indicated axes system,
assuming that Cd replaces the black Ti atom.}
\end{figure}

The first-principles calculations were performed with the WIEN97
implementation of the Full-Potential
Linearized-Augmented-Plane-Wave (FLAPW) method \cite{wien97},
which has been successfully used to determine the EFG in many
compounds \cite{succes_EFG,cdte_blaha}.
We worked in LDA approximation \cite{LDA}.
    We have considered a $2\times2\times3$ supercell
     of 72 atoms
   consisting of twelve
   unit cells of TiO$_2$ (see Fig.~\ref{fig1}),
where one of the 24 titanium atoms is replaced by a Cd atom.
   Each Cd atom is therefore almost equidistant from its Cd images
   at a distance of around 9 \AA.
   Substitution of a Ti  by a Cd atom  makes the resulting system
   metallic because of the lack of two electrons to fill up
   the $p$-O band.
   It is a question of central importance whether the real
   system we want to describe provides the lacking
    electrons (via an oxygen vacancy, for example) or not. In this letter
   we present the results of calculations assuming $(a)$ that
    extra electrons are not available (neutral impurity state: Cd$^0$),
   and  $(b)$ that the
   system somehow provides  the lacking two electrons (charged
   impurity state: Cd$^-$). In the situation
   $(a)$ we use the described SC performing calculations
   on a metallic system. To describe situation $(b)$ we add
   two electrons to the
   SC that we compensate with an homogeneous positive
   background in order to have a neutral cell to
   compute total energy and forces (\carg). We have also simulated situation $(b)$
   by replacing   the two most distant oxygen atoms from Cd by two fluorine atoms in order
   to provide the lacking two electrons
   without introducing any artificial background (Cd$^-$(fluorine)).

The value of the parameter RK$_{MAX}$ \cite{Soldner}, which
controls the size of the basis-set, was set to 6 and we introduced
local orbitals (LO) to include Ti-3$s$ and 3$p$, O-2$s$ and
Cd-4$p$ orbitals. Integrations in reciprocal space were performed
using the tetrahedron method taking
 8 k-points
in the irreducible Brillouin zone  for the metallic system $(a)$,
and 2 k-points for the non-metallic ones $(b)$. Once
self-consistency of the potential was achieved, the
  forces were computed 
  \cite{forces}, the ions were
displaced 
  \cite{newton}, and
new positions for Cd neighbors were obtained. The procedure was
repeated until the forces on the ions were below
 \mbox{0.025 eV/\AA}. At the relaxed structure, the
$V_{ij}$ elements of the EFG tensor were obtained
from the $V_{2M}$ components of
 the lattice
harmonic expansion of the self-consistent potential \cite{efg}.
  We have assumed that relaxations
  preserve the point-group symmetry of the SC, which
  restricts O1 and O2 displacements to {\bf YZ} plane and {\bf X} axis
  direction, respectively (see Fig.~\ref{fig1}). In order to check the
  stability of the solution obtained, at the end of the
  relaxation process we performed new calculations with O1 and O2
  atoms displaced from their symmetry positions and verified that
  this solution is a minimum.

\begin{figure}[!]
\includegraphics*[bb= 2.3cm 19.4cm 20cm 20cm, viewport=-1cm 0cm 10cm 3cm, scale=0.80]{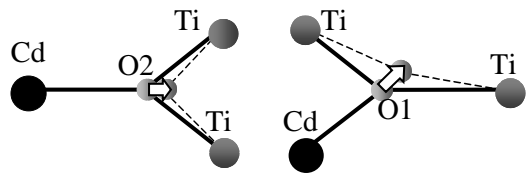}
 \caption{\label{fig2}
 Planes {\bf XZ} and {\bf YZ} (see Fig.~\ref{fig1}) containing O2 and O1
 atoms, respectively, with their neighbors. The arrows indicate the
displacement of the oxygen atoms from the unrelaxed  to the final
relaxed positions. The size of the relaxation has been duplicated
in order to better visualize the effect. }
\end{figure}

\begin{table*}
\caption{\label{tab:teor1} Distances from the Cd impurity to its
nearest neighbors (in \AA) and the EFG tensor principal components
at Cd site, $V_{ii}$(in \unitefg), for the relaxed structures of
the different systems considered in our calculation compared with
experiments and the  calculation of Sato {\it et al.}
 $\eta = (V_{22}-V_{11})/V_{33}$ ($|V_{33}|>|V_{22}|>|V_{11}|$). In the last row
distances and EFG tensor refer to Ti site in pure TiO$_2$. Q=0.83
b(Q=0.24 b) was used to calculate \cgav\ from the experimental
quadrupole coupling constant $\nu _Q$  at $^{111}$Cd($^{49}$Ti)
sites.}

\begin{ruledtabular}
\begin{tabular}{lcccccccc}
            & d(Cd-O1)  & d(Cd-O2) &  $V_{XX}$ & $V_{YY}$ & $V_{ZZ}$
            & \cgav & \cgav-direction & $\eta$    \\
\hline
 \neutra\    &  2.15    &   2.11  &  -7.16    & +6.82    &  +0.34   &  -7.16  &  \textbf{X}   & 0.91     \\
 \carg\     &  2.18    &   2.11  &   -2.87    & +4.55    &  -1.68   &  +4.55  & \textbf{Y} & 0.26     \\
 \fluor\    &  2.19    &   2.12  &   -2.46    & +4.10    &  -1.63   &  +4.10  & \textbf{Y} &0.20     \\
 Exp.
  (singlecrystal)     &     &    &    &     &       &   5.34(1) &  \textbf{X} or \textbf{Y}  & 0.18(1) \\
 Exp.
  (polycrystal)\cite{lieb,catchen} &     &     &     &     &     &   5.23(5)/5.34(2)  &  ---   & 0.18(1)/0.18(1)\\
 Calc.\cite{akai}    &  2.06    &   2.10  &   +1.54    & +3.56    &  -5.09   &  -5.09  & \textbf{Z} &0.39     \\
\hline
 Exp.
  (pure TiO$_2$)\cite{nmr_tio2}     &  1.94    &   1.98  &     &    &    &  2.2(1)  & \textbf{Z} &0.19(1)     \\
\end{tabular}
\end{ruledtabular}
\end{table*}
In Table~\ref{tab:teor1} we show the results for the relaxation of
the 6  nearest oxygen neighbors of the Cd impurity. We see that
for both impurity charge-states the relaxations are quite
anisotropic, with the Cd-O1 distance bigger than the Cd-O2
distance, opposite to the initial unrelaxed structure. This
conclusion contradicts the assumptions of previous studies of this
system \cite{akai,lieb} and confirms the tendency predicted in the
first calculation performed by some of us with a much smaller SC
\cite{our_nqi}. Anisotropy in the relaxations of the nearest
oxygen neighbors of the Cd-impurity can be understood by
inspection of Fig.~\ref{fig2}.
Stretching of Cd-O2 bond implies a considerable shortening in
Ti-O2 bonds. However, stretching of Cd-O1 bond affects the Ti-O1
bonds not so much, since the structure is more open in this
direction. So, at the end of the relaxation, the Cd-O1 bond
stretches almost twice as much as the Cd-O2 bond. Relaxations and
EFGs are very similar for the two ways of simulating the charged
state of the impurity indicating that both approaches are well
suited to deal with this problem. These results for EFG agree
reasonably well with available experimental information
\cite{lieb,catchen} and are very different from the ones obtained
for the neutral state of the impurity, \neutra. The huge
difference obtained in the $V_{ii}$ components for the two
impurity charge-states is mainly due to the filling of the
impurity state at the Fermi level which has an important component
of \cddyz\ character \cite{PRB_nos}. The high $\eta$ value
obtained for \neutra\ is too far from experimental findings
suggesting that Cd as an impurity in TiO$_2$ is in a charged
state. The results obtained for \carg\ changes very slightly when
the coordinates of atoms beyond the Cd-nearest neighbors are
relaxed (see Table~\ref{tab:teor2}).

\begin{table}
\caption{\label{tab:teor2} Results of the different relaxations
performed for system \carg. $N_A$ is the number of atoms (within a
radius $R_C$ from Cd) that relax in each case. All units as in
Table~\ref{tab:teor1}.}
\begin{ruledtabular}
\begin{tabular}{lrcccccc}
           $R_C$ & $N_A$ & d(Cd-O1)   & d(Cd-O2)  &
                  $V_{XX}$ & $V_{YY}$  & $V_{ZZ}$ & $\eta$  \\
                  \hline
 2.5   &  6    &  2.185   &  2.111    &
                 -2.87    & +4.55     &  -1.68   &  0.26    \\
 4.0   &  24   &  2.176    &   2.104  &
                   -3.25  & +4.99     & -1.74  & 0.30         \\
 4.6   &  42  &  2.187   &   2.116    &
                   -3.17   & +4.86   &   -1.69    &  0.30     \\
\end{tabular}
\end{ruledtabular}
\end{table}

 In addition we also performed FLAPW calculations for the system \carg\
imposing the constraint of isotropic relaxations of O1 and O2
atoms and obtained \cgav\ pointing in [001]-direction and a high
$\eta$ value of 0.91 confirming for a converged-size SC that an
isotropic relaxation is not consistent with the available
experimental data.

 To check the accuracy and convergence of our results we
have performed additional
  calculations at the
relaxed structure of case \carg\ with $R_C$=2.5 {\AA} considering
$RK_{MAX}$=7, including LO for Cd-$4d$ orbitals, increasing the
mesh in $\vec k$-space, using GGA approximation \cite{gga} for
exchange-correlation potential and also considering
 $2 \times 2 \times 2$ and  $2 \times 2 \times 4$
 SCs. These checks show that  d(Cd-O1)  and d(Cd-O2)
 are converged within 0.01 \AA and \cgav\ and $\eta$ within 0.5 \unitefg\
 and 0.1, respectively. Details will be presented in a forthcoming
  paper \cite{PRB_nos}.


In order to check  the change in orientation of \cgav\ from the
{\bf Z} to the {\bf Y} direction upon substitution of a Ti atom by
a Cd impurity, we performed  PAC experiments in TiO$_2$ single
crystals with the $^{111}$Cd probe.
 The PAC technique measures the probability
W($\mathbf{k}_1$,$\mathbf{k}_2$,t) to detect the second $\gamma
$-ray of a nuclear-decay $\gamma $-$\gamma $ cascade at the
direction $\mathbf{k}_2$($\theta_2$,$\phi_2$) after a time $t$ the
first $\gamma $-ray was detected at the direction
$\mathbf{k}_1$($\theta_1$,$\phi_1$), where $\theta_i$ and $\phi_i$
describe the propagation direction of $\gamma_1$  and $\gamma_2$
with respect to a chosen quantization axis.
The angular correlation may be modulated in time by hyperfine
interactions which induce transitions between the m-states of the
intermediate level of the cascade, changing their population
distribution (relative to $\gamma_1$) and thus the angular
distribution of $\gamma_2$. The correlation function
W($\mathbf{k}_1$,$\mathbf{k}_2$,t) therefore carries information
on the strength, symmetry and time dependence of the hyperfine
field and is sensitive to the orientation of the interaction with
respect to $\mathbf{k}_1$ and  $\mathbf{k}_2$.

The orientation dependence is easily understood by considering two
simple cases. Let us assume an axially symmetric EFG  and choose
the direction of $\mathbf{k}_1$ as quantization axis. If we orient
the EFG parallel to $\mathbf{k}_1$, the spin precession does not
change the projection of $\textit{I}$ on $\mathbf{k}_1$. As a
consequence the population of the $\textit{m}$-states with respect
to $\mathbf{k}_1$ and thus  the correlation function remain
constant in time. In case the symmetry axis of the EFG is
perpendicular to the direction of the first ($\mathbf{k}_1$)
detector, the spin precession causes a periodic oscillation of the
projections of $\textit{I}$ on $\mathbf{k}_1$, corresponding to
transitions between the different $m$-states relative to
$\mathbf{k}_1$, and a periodic time-modulation of the intensity
seen by the second detector results, which is invariant under
rotation around the symmetry axis of the interaction. Details of
the PAC theory
can be found in Ref.~\cite{TDPAC}.

For the  PAC experiment, we used a commercially available TiO$_2$
single crystal with  the dimensions  10x10x0.5 mm. The crystal was
cut with the $[001]$ axis perpendicular to the 10x10 mm faces and
the $[110]$ axis perpendicular to the 10x0.5 mm-sides. The mother
isotope $^{111}$In of the $^{111}$Cd PAC probe  was implanted at
room temperature into the 10x10 mm faces  with an energy of 160
keV to a dose of $\approx$10$^{13}$ ions/cm$^2$. Following the
implantation the single crystal
 was annealed at  800$^\circ$C
for 4 h in order to remove any radiation damage. After this
treatment at least 90$\%$ of the probes were found to reside on
the regular Ti site.
\begin{figure}[h]
\includegraphics*[bb= 41 454 10 10, viewport=0cm -0.02cm 8.5cm 5.6cm, scale=0.86]{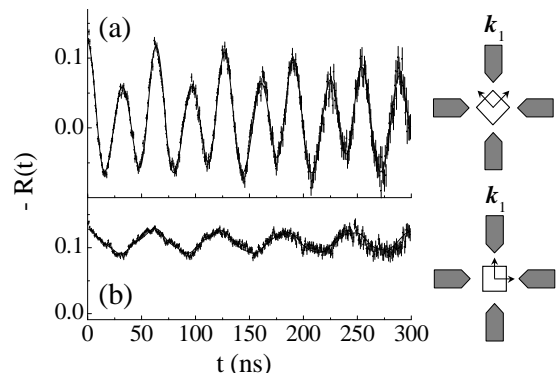}
%
%
\caption{\label{fig3} PAC  spectra of the
$^{111}$In/$^{111}$Cd-implanted TiO$_2$ single crystal with (a)
the $[110]$ axis pointing at 45$^{\circ }$ from detectors; (b) the
$[110]$ axis pointing to one detector.}
\end{figure}

The time-differential perturbed-angular correlation of the 175 and
247 keV $\gamma$-$\gamma$ transition of the cascade $^{111}$Cd was
measured with a standard four  BaF$_2$ detectors PAC spectrometer
for several orientations of the crystal relative to the detectors.
The spectra shown in Figs.~\ref{fig3}(a) and \ref{fig3}(b) were
obtained with the $[001]$-axis normal to the detector plane. In
the case of Fig.~\ref{fig3}(a) the $[110]$-axis was parallel to
the bisector of two detectors, in fig. 3b it was parallel to one
detector. The visual inspection of these spectra already reveals
the EFG orientation:

  Clearly, the rotation around the $[001]$ direction causes
substantial changes of  the PAC pattern. This observation
immediately  discards the $[001]$ direction as the direction of
V$_{33}$  because, as  pointed out above, in the case of an
axially symmetric EFG  the PAC pattern should be invariant under
rotation around the symmetry axis of the interaction. In the
present case the EFG presents a slight axial asymmetry ($\eta =
0.18$). But even then, the changes upon rotation around the
V$_{33}$ axis normal to the detector plane should be imperceptibly
small. This is easily checked by calculating the perturbation
factor for this geometry.

If in a crystal of tetragonal symmetry such as TiO$_2$ the EFG is
not parallel to the $[001]$ axis, one must  expect at least two
identical EFG tensors with equal weight but different orientations
of the principal axis system. The PAC pattern obtained with the
$[110]$-axis at 45$^\circ$ to the detectors (Fig.~\ref{fig3}(a))
can be described with a single site, i.e., all $^{111}$Cd probes
are subject to the same EFG implying that the two expected EFG
orientations must be equivalent. With the $[110]$-axis along the
bisector of the detectors, two equivalent orientations can be
obtained only with V$_{33}$ in the basal plane parallel to either
the $[110]$- or $[1\overline{1}0]$-axis.

A fit of the exact theory \cite{barradas} to the spectrum of
Fig.~\ref{fig3}(a) gave the values $\theta \leq 8 ^\circ$  and
 $\phi$ = 43(2)$^\circ$ for the angle  between
V$_{33}$ and the detector plane and the angle between
$\mathbf{k}_1$ and the projection of V$_{33}$ onto the detector
plane, respectively. We estimate that we adjusted the 10x10mm face
of the TiO$_2$ single crystal parallel to the detector plane
within $\pm 5 ^\circ$. So, the value of $\theta = 0^\circ$ is
compatible with the experimental accuracy.

The conclusion of V$_{33}$ lying in the basal plane parallel to
either the $[110]$- or $[1\overline{1}0]$-axis is confirmed by the
spectrum in Fig.~\ref{fig3}(b). In this case the two EFG
orientations are not equivalent. As discussed above, with the EFG
parallel to the $\mathbf{k}_1$ detector, the angular correlation
remains unperturbed while the EFG normal to $\mathbf{k}_1$
produces a periodic modulation. The superposition of these two
configurations leads to a PAC pattern consisting of a
small-amplitude oscillation on a constant background. The
oscillation is mainly caused by the second configuration, but the
exact theory shows that a small oscillatory contribution also
comes from the slight asymmetry of the  EFG. The least-squares fit
to the spectrum in Fig.~\ref{fig3}(b) gave the angles
 $\phi_1 = 3(1)^\circ$ and $\phi_2 = 90(1)^\circ$. The value of
the angle $\theta$ is consistent with the one obtained from the
spectrum in Fig.~\ref{fig3}(a). For both spectra the fitted values
$\nu _Q = 107.1(3)$ MHz and   $\eta = 0.18(1)$ are in good
agreement with those previously reported \cite{lieb,catchen}.

In summary, we have shown that the inclusion of Cd-impurities in
TiO$_2$ distorts the oxygen octahedron surrounding the impurity,
producing a change in the EFG tensor orientation. Such a change
does not result if an isotropic relaxation of the oxygen
octahedron is assumed. According to our calculation, the asymmetry
of the EFG tensor suggests that Cd in TiO$_2$ is not neutral, but
in a charged state. We therefore conclude that a proper
theoretical approach of electronic properties of metal impurities
in oxide semiconductors should consider the charge state of the
impurity and the impurity-induced relaxations without constraints
in a fully self-consistent way.

This work was
 supported by CONICET, ANPCyT, F.
Antorchas (Argentina), and DFG (Germany). MR is indebted to J-P.
Dallas (CECM, Vitry) for the single crystals orientation and to A.
Traverse (LURE) for valuable support on this project. We
thank R. Vianden (ISKP) and A.F. Pasquevich (UNLP) for  critical
comments.


\newpage

\end{document}